\documentclass[fleqn,10pt]{SelfArx} 

\usepackage{lipsum} 

\definecolor{color1}{RGB}{0,0,90} 
\definecolor{color2}{RGB}{0,20,20} 

\usepackage{hyperref} 
\hypersetup{hidelinks,colorlinks,breaklinks=true,urlcolor=color2,citecolor=color1,linkcolor=color1,bookmarksopen=false,pdftitle={Title},pdfauthor={Author}}

\JournalInfo{$ $}
\Archive{$ $}
\PaperTitle{Relaxation algorithm in description of superconducting structures. }

\Authors{Krzysztof Pomorski *\textsuperscript{1}\textsuperscript{2} , Przemyslaw Prokopow\textsuperscript{3} , Mariusz Zubert\textsuperscript{4}}
\affiliation{\textsuperscript{1}\textit{Faculty of Physics, Astronomy and Applied Computer Science , Jagiellonian University, Cracow, Poland}}
\affiliation{\textsuperscript{2}\textit{Faculty of Physics, University of Warsaw, Warsaw, Poland}}
\affiliation{\textsuperscript{3}\textit{RIKEN, Japan}}
\affiliation{\textsuperscript{4}\textit{Department of Microelectronics and Computer Science, Lodz University of Technology, Lodz, Poland}}
\affiliation{*\textbf{E-mail address}: kdvpomorski@gmail.com}
\newcommand{\keywordname}{Keywords:relation algorithm, superconducting structures}

\Abstract{Relaxation method is described on simple superconducting cases in one and two dimensions in Ginzburg-Landau (Gl) formalism.
The structure of the algorithm is given for the case time independent and time dependent GL equation. The advantages and disadvantages of the algorithm
are specified. Particular cases solved by the relaxation algorithm is unconventional Josephson junction (uJJ), modified uJJ, SQUID built on the base of uJJ
and superconduting current limiter. The artifacts of relaxation algorithm are described.}

\begin{document}

\flushbottom 

\maketitle 

\tableofcontents 

\thispagestyle{empty} 

\section*{Introduction to the relaxation algorithm} 
Variational calculus is used in many branches of fundamental and applied science. In theoretical mechanics the minimization of action principle is used and it allows to derive the equation of motion. Maxwell equations can also be derived using variational derivative on action with respect to physical quantities as electric or vector potential. Also quantum mechanics and theory of relativity can be formulated by minimization of action function and hence
variational derivative with respect to desirable quantities will result in proper equations of motion. Refined and theoretical review  of relaxation method is given by Adler \cite{AdlerRMP}.
In this work we will consider only on effective formulation of superconductivity problems, which are similar in many ways to theory of superfluidity.
We will concentrate on practical aspects of application of relaxation method. In Ginzburg-Landau (GL) theory we have free energy functional and
equations of motion are given by
\begin{equation}
    \label{simple_equation}
    \frac{\delta F[X_{i,j,k,...}]}{\delta X_{i,j,k,...}} =0, 
\end{equation}
where $X_{i,j,k,...}$ is the physical field e.g. A-vector potential. More details on derivation of GL equations can be found in \cite{GLderivation}.
In numerical computations we adapt the scheme
\begin{equation}
    \label{simple_equation}
    \frac{\delta F[X_{i,j,k,...}]}{\delta X_{i,j,k,...}} =- \eta_{i,j,k,...} \frac{\Delta X_{i,j,k,...}}{\Delta t_{i,j,k,...}},
\end{equation}
where $\Delta t_{i,j,k,...}$ and $\eta_{i,j,k,...}$ is kept constant, while $\Delta X_{i,j,k,...}$ is changing during the simulation.
Using GL free energy functional $F[\psi,A]$ we obtain the Ginzburg-Landau equations
\begin{equation}
    \label{simple_equation}
    \frac{\delta F[\psi,A]}{\delta \psi(x)} =- \eta_1 \frac{\Delta \psi(x)}{\Delta t_1} ,  \frac{\delta F[\psi,A]}{\delta A} =- \eta_2 \frac{\Delta A}{\Delta t_2}.
\end{equation}
At first we consider the simple one dimensional form of Ginzburg-Landau equation of the form
\begin{eqnarray} 
(\alpha(x,t)+\beta(x)|\psi(x,t)|^2+\frac{1}{2m}(\frac{\hbar}{i}\frac{d}{dx} \nonumber \\
-\frac{2e}{c}A_x(x,t))^2-\gamma \frac{d}{dt})\psi(x,t)=0, \label{1dimGLequation} 
\end{eqnarray} 
where $\gamma$ is some constant. Maxwell equation gives
\begin{equation}
\nabla \times (\nabla \times A))= \mu_0 j_{curr}(t) +\mu_0\epsilon_0  \frac{\partial E(t)}{\partial t},
\end{equation}
where $j$ is the electric current density from GL theory. The superconductor-vacuum and superconductor-nonsuperconducor interface in yz plane is accounted
in GL theory by condition $(\frac{\hbar}{i}\frac{d}{dx}-\frac{2e}{c}A_x(x))\psi(x)=0$ or $(\frac{\hbar}{i}\frac{d}{dx}-\frac{2e}{c}A_x(x))\psi(x)=\frac{1}{b}\psi(x)$, where b is some material depending on superconductor and non-superconducting
material. For certain type of problems with translational symmetry in z direction and only $A_z$ component we can adapt the gauge $\nabla A(x,y)=0$.
Let us limit to one dimensional problems. Since we deal with numerical algorithm we obtain the approximate values of $\psi(x,t)$ across given lattice. Thus it will be helpful to introduce space error function for GL equation given as $e(x)$

\begin{equation}
e(x)=\frac{|(-\frac{\hbar^2}{2m}(\frac{\hbar}{i}\frac{d}{dx}-\frac{2e}{c}A_x(x))^2+\alpha(x)+\beta(x)|\psi(x)|^2) \psi(x))|}{|\psi(x)|+|\psi_{min}|},
\end{equation}
where $|\psi_{min}|$ is small constant.
Since relaxtion method rely on many iterations with time it is also useful to introduce average error $e_{av}$ defined as
$e_{av}=\frac{\sum_{i=1}^{N_x} e(x_i)}{N_x}$. It is quite straightforward to generalize $e(x)$ and $e_{av}$ for two and three dimensions.

\addcontentsline{toc}{section}{Introduction to the relaxation algorithm} 

\section*{Implementation of the relaxation algorithm}

\addcontentsline{toc}{section}{Implementation of the relaxation algorithm.}

\begin{figure*}[ht]\centering 
\includegraphics[width=\linewidth]{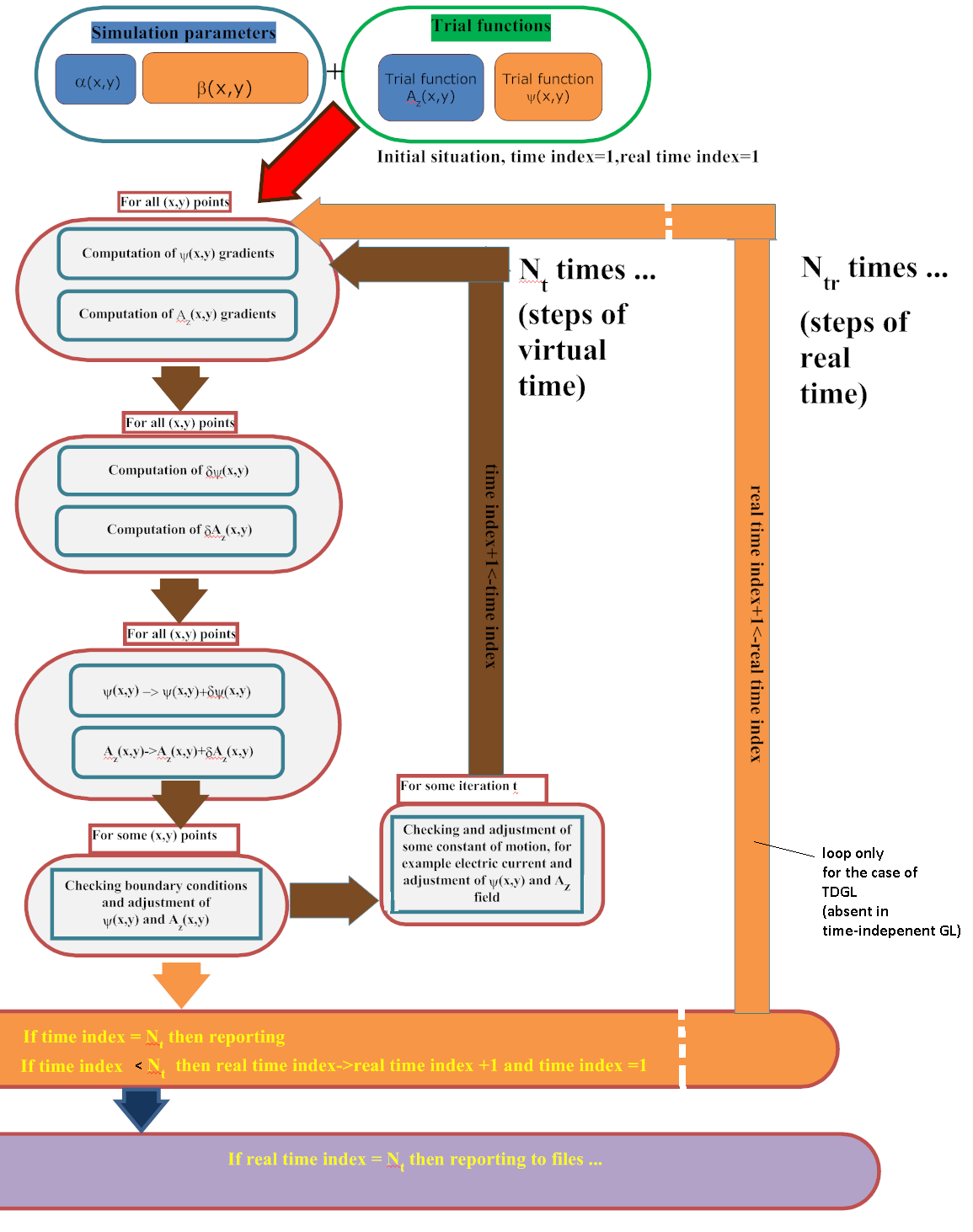}
\caption{Schematic description of relaxation algorithm solving $GL/TDGL$ equations.}
\label{TDGL_scheme}
\end{figure*}

Let us apply the relaxation method to the one dimensional GL equation \ref{1dimGLequation}, which is schematically depicted in Fig.\ref{TDGL_scheme}.
At first we make initial guess of vector potential $A_x(x)$ and $\psi(x)$. We chose certain lattice of
$N_x$ points and values of $\eta_1$, $\eta_2$, $\Delta t_1$ and $\Delta t_2$. We compute initial electric current density $j_{curr}$ from GL theory and gradients of $\psi$ and $A_x(x)$.
Next step is the execution of  $N_t$ times the loop:
\\
\\
1. We compute  $j_{curr}$ on every element of space lattice. \\
2. We compute $\frac{d}{dx}\psi(x,t)$, $\frac{d^2}{dx^2}\psi(x,t)$, $\frac{d}{dx}A_x(x,t)$ and $\frac{d^2}{dx^2}A_x(x,t)$.
for each element of lattice. \\ 
3. We compute the change of vector potential and $\psi$ for every element of lattice
\[  \left\{
  \begin{array}{l l}
    \Delta A_x(x)=- \Delta t_1 \eta_1 j_{curr}(x),  \\
    \Delta \psi(x)=\Delta t \eta (\alpha  + \beta |\psi(x)|^2 +\frac{1}{2m}(\frac{\hbar}{i}\frac{d}{dx}-\frac{2e}{c}A_x(x))^2)\psi(x) 
  \end{array} \right.\]
\\
4.  We apply changes of $\Delta \psi$ and $\Delta A_x(x)$ for each element of lattice \\
 \[  \left\{
  \begin{array}{l l}
    \psi(x) \rightarrow \psi(x)+\Delta \psi(x),  \\$ $
    A_z(x,y) \rightarrow A_x(x)+\Delta A_x(x).
  \end{array} \right.\]
5. We check the correctness of boundary conditions and make the adjustments in $\psi$ so they are fulfilled.  \\
6. We make the correctness of certain physical constrains and make the necessary adjustments in $\psi$ and $A_x$ field. \\
7. We compute free energy F, numerical error e(x) and $e_{av}$.
\\
\\
The criteria necessary for obtaining the solution is the minimization and saturation of numerical error $e(x)$, $e_{av}$ and free energy F.

\section*{Various solutions obtained by the relaxation algorithm}

\addcontentsline{toc}{section}{Various solutions obtained by the relaxation algorithm}

\begin{figure} 
\includegraphics[width=0.9\linewidth]{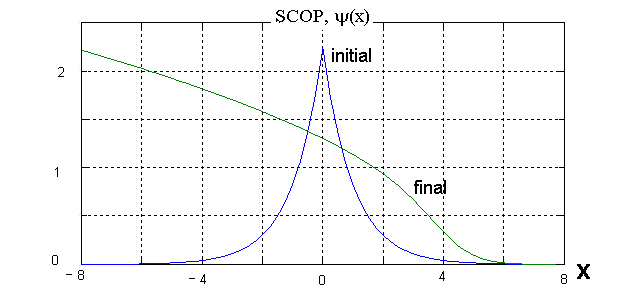}
\caption{Distribution of superconducting order parameter (SCOP) $\psi(x)$ before and after simulation for the case of 1 dimensional superconductor. }
\label{fig:view1}
\end{figure}

\begin{figure}
\includegraphics[width=0.9\linewidth]{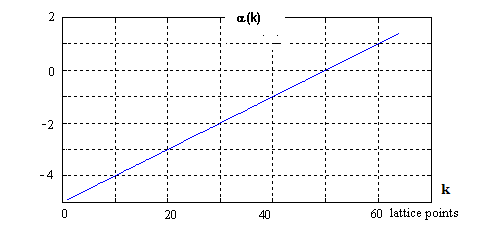}
\caption{Distribution of $\alpha$ coefficient. }
\label{fig:view5}
\end{figure}

\begin{figure}
\includegraphics[width=0.9\linewidth]{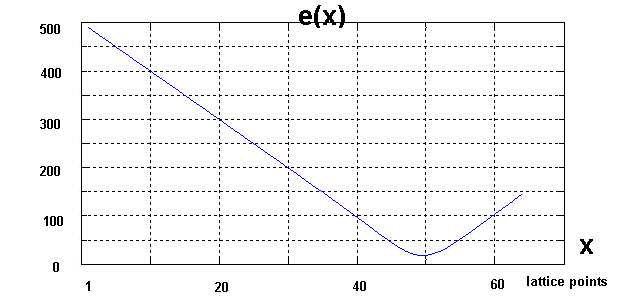}
\caption{Distribution of relative error e(x) obtained in final step of simulation.}
\label{fig:view6}
\includegraphics[width=0.9\linewidth]{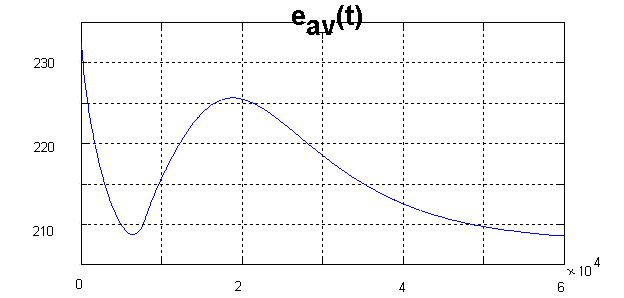} 
\caption{Evolution of average error $e_{av}(t)$ with simulation steps.}
\label{fig:view7}
\end{figure}

\begin{figure}
\includegraphics[width=0.9\linewidth]{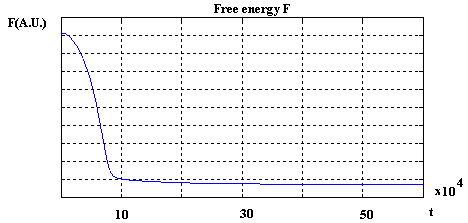}
\caption{Free energy F vs iteration t.}
\label{fig:view8}
\end{figure}

At first one dimensional GL equation given in \ref{1dimGLequation} was solved with use of relaxation algorithm for the case of zero vector potential and $\alpha$ depicted in Fig.\ref{fig:view5}. The initial and final values of $\psi$ are given in Fig.\ref{fig:view1}.
The error relative error e(x) in the last step of simulation is depicted in Fig.\ref{fig:view6}. The final solution is obtained after the free energy $F$ and
average error $e_{av}$ are minimizing and get the saturation what is depicted in Fig.\ref{fig:view7} and in Fig.\ref{fig:view8}.
Despite the fact that initial values of $\psi$ were far from physical intuition proper numerical solution (corresponding to physical intuition) was obtained.

Next example was solving two dimensional GL equation of the form
\begin{figure*}[ht]\centering 
\includegraphics[width=0.4\linewidth]{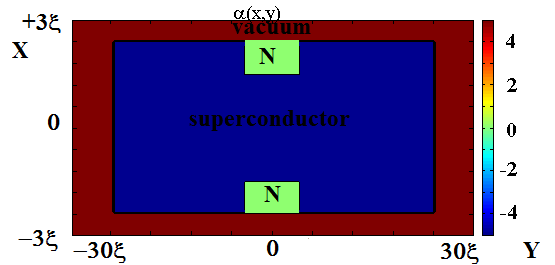} \includegraphics[width=0.4\linewidth]{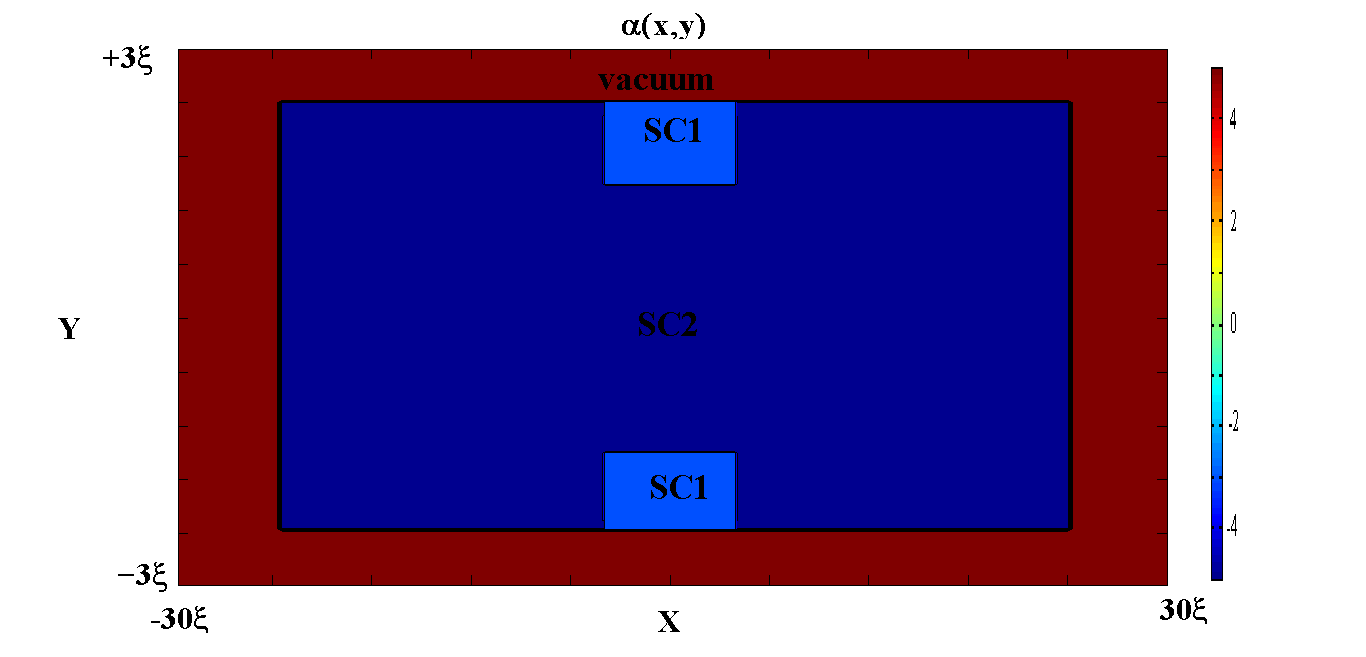}
\caption{Distribution of $\alpha(x,y)$ defining modified unconventional Josephson junction (left side) and half modified unconventional Josephson
junction (right side).}
\label{fig:view3}
\includegraphics[width=0.4\linewidth]{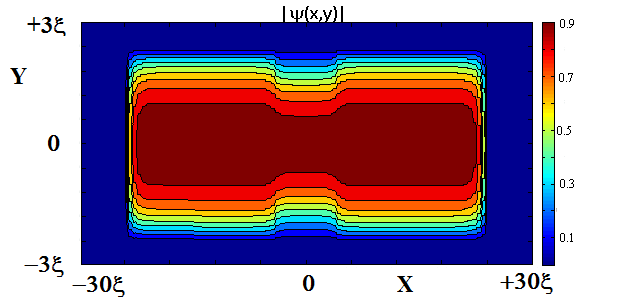}\includegraphics[width=0.4\linewidth]{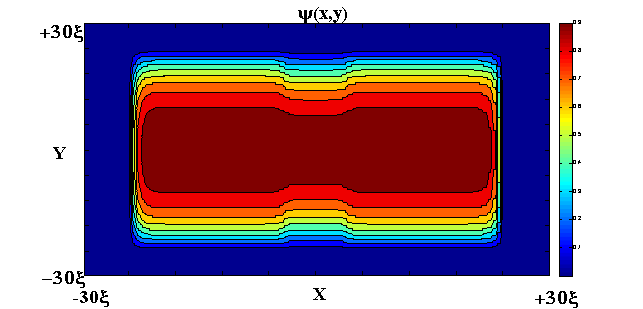}
\caption{Distribution $\psi(x,y)$ in modified (left side) and half modified (right side) unconventional Josephson junction obtained by the relaxation method.}
\label{fig:view2}
\end{figure*}

\begin{figure*}[ht]\centering 
\includegraphics[width=0.4\linewidth]{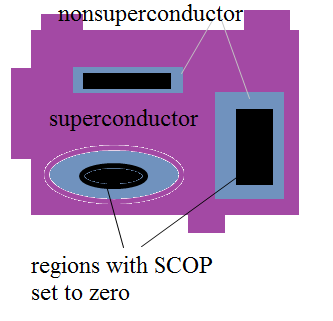}\includegraphics[width=0.4\linewidth]{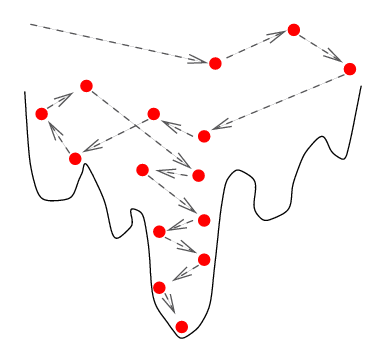}
\caption{The introduction of stabilizing regions, which has the imposed $\psi(x,y)=0$ (left side). In order to confirm the stability of solutions the
numerical noise is introduced to the relaxation algorithm. It helps to achieve the global minima for F functional as depicted on the right picture.
Such situation is analogical to the case of random dropping balls in random hills potential.}
\label{fig:view4}
\end{figure*}

\begin{eqnarray}
-\frac{\hbar^2}{2m}(\frac{d^2}{dx^2}+\frac{d^2}{dy^2}) \psi(x,y)+ (\frac{\hbar^2}{2m}A^2_z(x,y)\frac{4e^2}{c^2}+ \nonumber \\ \alpha(x,y)+\beta(x,y)|\psi(x,y)|^2)\psi(x,y)=0.
\end{eqnarray}

At first we consider the case of zero magnetic field what implies $A_z(x,y)=0$.
We continue the study on unconventional Josephson junction started by \cite{KPPSSB} when we place the non-superconducting element on the top of
superconductor slab. This time the non-superconductor strip is placed inside the superconductor. We also can consider placement of less superconducting
bar on the more superconducting bar (more negative $\alpha$ coefficient). Such structures are depicted on the left and right side of Fig.\ref{fig:view3}, which are
given by certain distribution of $\alpha(x,y)$ field.

The presence of normal strip reduces superconducting order parameter (SCOP) in more significant way as in the case of weakening
of the superconductor what is given on the left and right side of Fig.\ref{fig:view2}.
The diffusion of Cooper pair from the superconducting strip into superconducting strip takes place.
In the case with two superconductors superconducting order parameter diffuses from more superconducting region into less superconducting region.

The situation with modeling superconducting mesoscopic structures becomes complicated when there is occurrence of non-zero vector potential and hence magnetic field.
We can measure the error of $\psi$ function as it was introduced by means of $e(x)$ function.
In order to measure the error of vector potential it is necessary to consider the equation for vector potential.
We have
\begin{equation}
\nabla^2 A_z(x,y)=-C A_z(x,y) |\psi(x,y)|^2,
\end{equation}
where C is constant quantity depending on fundamental physical constants.
Quite obviously from the last equation
\begin{equation}
C(x,y)=\frac{\nabla^2 A_z(x,y)}{A_z(x,y) |\psi(x,y)|^2 }
\end{equation}
one should obtain the constant value C. In real numerical simulation C values would change and be position dependent as C(i,j) for the
case of discrete lattice.
Therefore the best criteria is the minimization of quantity $C_1$ is defined (seems to be defined) as
\begin{equation}
C_1=\sum_{k,l} \frac{(|\frac{d}{dx}C(k,l)|+|\frac{d}{dy}C(k,l)|)}{|C(k,l)|+|C_{min}|},
\end{equation}
where summation is conducted over the all points of lattice $(N_x,N_y)$ and $C_{min}=10^{-3}$ for example. One could define many similar functions as $C_1$ that should be minimized in the iterations of the relaxation algorithm.
Quite much similar reasoning could be presented in case of 3 dimensions.
All considerations are valid for the case of time independent and time dependent GL equations.

\section*{Artifacts and limitations of the relaxation method}

\addcontentsline{toc}{section}{Artifacts and limitations of the relaxation method}
Relaxation method is quite stable especially when iteration step $\Delta t$ is small. For obvious reasons it cannot be too small since we expect the simulation to be finished in reasonable time. Nevertheless sometimes it is necessary to stabilize its output. One tested method is by keeping certain regions of the lattice with value of superconducting order parameter set to zero as it is depicted in Fig.\ref{fig:view4}.
In order to confirm the stability of solution is the addition of noise to the system. After certain time the system should recover
all values of $A_z(x,y)$ from before the noise addition. In this way we can become more sure that we have obtained the stable numerical solution.
This is important since the space of initial probe as $A_z(x,y)$, $\psi(x,y)$ functions is infinite. In general from the observation we have noticed that the probe function should have bigger monotonicity change than the expected numerical solution.
We have tested the relaxation algorithm for the case of superconducting rectangular for d-wave superconductor in ab-plane.
Topology of solutions of GL$(x^2-y^2)$ obtained by the relaxation method was in accordance with solutions obtained by different methods as described by \cite{DwaveSquare}.

\section*{Further perspectives} 
The relaxation method used for study of superconducting mesoscopic structures is very stable even in the cases of more complex GL functionals as given in \cite{KPPSSB}. This method have the capacity to model very complex mesoscopic superconducting structures.
The parallelization of the relaxation method should be introduced. Because of its simplicity we have the reasons to believe that relaxation method could be the core for building
universal platform capable of modeling many types of superconducting devices and various mesoscopic structures.
Various results obtain in \cite{GLNumericalEffect} needs to be further confirmed by the relaxation method.
\addcontentsline{toc}{section}{Further perspectives}


\section*{Acknowledgments and References} 

\addcontentsline{toc}{section}{Acknowledgments} 
We would like to thank to A.Majhoffer, A.Bednorz \\
and J.Lawrynowicz for helpful discussions. 
\section*{References} 
\addcontentsline{toc}{section}{References} 

\begin{enumerate}
  \item Stephen L. Adler and Tsvi Piran. Relaxation methods for gauge field equilibrium equations. Rev. Mod. Phys., 56:1–40, Jan 1984. 
  \item A.V.Dmitriev and W.Nolting. On details of the thermodynamical derivation of the Ginsburg–Landau equations, arXiv, 0312094, 2003.
  \item K.Pomorski and P.Prokopow. Possible existence of field induced Josephson junctions. Physica Status Solidi B, 249:1805–1813, 2012. 
  \item G.Haran and P.Pisarski. D-wave superconductivity in a system with open boundary conditions. Acta Physica Polonica A, 2004. 
  \item W.B.Richardson et ac. Numerical effects in the simulation of Ginzburg–Landau models for superconductivity. International Journal for numerical methods in engineering, 2004.
\end{enumerate}

\end{document}